\documentclass[journal]{IEEEtran}
\usepackage{siunitx}
\usepackage{xcolor,soul,framed} 

\colorlet{shadecolor}{yellow}
\usepackage[pdftex]{graphicx}
\graphicspath{{../pdf/}{../jpeg/}}
\DeclareGraphicsExtensions{.pdf,.jpeg,.png}

\usepackage[cmex10]{amsmath}
\usepackage{array}
\usepackage{mdwmath}
\usepackage{mdwtab}
\usepackage{eqparbox}
\usepackage{url}
\usepackage{comment}

\begin{document}
\bstctlcite{IEEEexample:BSTcontrol}
    \title{Electromagnetically Transmissive and Thermally Opaque Engineered Microwave Interconnect}
  \author{Md.~Jawad~Ul~Islam,~\IEEEmembership{Graduate~Student~Member,~IEEE},
Ishrat~Jahan, 
Abdelrahman~Omar,~\IEEEmembership{Graduate~Student~Member,~IEEE},
Md.~Omar~Faruk~Noman,
Md~Zunaid~Baten,~\IEEEmembership{Senior~Member,~IEEE},
Soumitra~R.~Joy,~\IEEEmembership{Member,~IEEE}

 \thanks{This work is supported by University of North Carolina at Charlotte Faculty Startup Grant.}

  \thanks{ M. J. U. Islam and M. Z. Baten are with Bangladesh University of Engineering and Technology, Dhaka, Bangladesh (e-mail: jawaduid@eee.buet.ac.bd; mdzunaid@eee.buet.ac.bd).}%
  \thanks{ I. Jahan, A. Omar, M. O. F. Noman and S. R. Joy are with University of North Carolina at Charlotte, NC, USA
  (e-mail: ijahan1@charlotte.edu; aomar5@uncc.edu; mnoman@charlotte.edu; s.joy@charlotte.edu).}
  } 

\markboth{
}{}

\maketitle

\begin{abstract}

In a standard two-conductor microwave transmission line, both the signal 
trace and the wide ground plane act as parasitic heat leakage paths along 
the propagation direction. To prevent this heat leakage, we devised a 
single-conductor microwave interconnect: the ground plane is removed 
entirely, and the remaining signal trace is periodically fragmented into 
physically isolated metallic islands. We show thato despite the 
galvanic isolation between adjacent metal islands, a surface-confined mode 
of specific symmetry, the odd-symmetric spoof surface plasmon polariton, 
propagates through this periodic, single-conductor medium. This odd symmetry imposes 
an electromagnetic field null at the waveguide centerline, allowing complete 
removal of the metal at periodic intervals along the single-conductor 
waveguide without impacting EM propagation. Measurements on fabricated prototypes confirm the transmission 
response despite the physical fragmentation of the conductor, with 
de-embedded propagation loss of $\sim$0.0031~dB/mm across a 5--6.5~GHz 
passband. Time-domain characterization verifies 1~Gbps amplitude-modulated 
data transmission with less than 145~ps group delay variation. Coupled 
electro-thermal simulation yields an effective thermal conductivity of 
0.336~W/m$\cdot$K, a 40$\times$ reduction relative to a continuous microstrip 
line of identical footprint.

\end{abstract}


%
\IEEEpeerreviewmaketitle


\section{Introduction}

\IEEEPARstart{M}{etallic} interconnects are the dominant signal carrier in 
microwave systems, but they also form an unavoidable thermal conduction path. 
The Wiedemann-Franz law~\cite{ashcroft1976} directly couples a metal's 
thermal conductivity to its electrical conductivity, so any continuous 
metallic signal trace simultaneously functions as a heat pipe. In systems 
where the signal path connects components operating at different temperatures 
— thermally sensitive receiver front ends, active phased-array 
transmit/receive modules, or circuits spanning a temperature 
gradient~\cite{gao2020apaa, xu2026apaa} — parasitic heat flow through the 
signal conductor degrades the performance of temperature-critical components.

Preventing this requires breaking the continuity of the metallic signal 
trace. While capacitive discontinuities in metallic traces are routinely introduced in 
bandpass filter designs~\cite{yin2016capacitive, Wang2019}, these structures 
all retain a continuous ground plane to support a transverse electromagnetic 
(TEM) guided mode. In a capacitively-coupled TEM line, displacement current 
bridges the gaps in the signal trace, but conduction current still flows 
uninterrupted through the physically wide ground plane, providing a parallel heat conduction 
path that negates any thermal benefit from the signal trace discontinuities. 
Achieving thermal isolation along a microwave signal path therefore imposes 
two simultaneous requirements: first, the ground return path must be 
eliminated entirely, departing from the textbook two-conductor transmission 
framework; second, the remaining single-conductor metallic structure must 
itself be galvanically fragmented to prevent thermal conduction along the signal 
direction.

\begin{figure}
\centering
\includegraphics[width=3.45in]{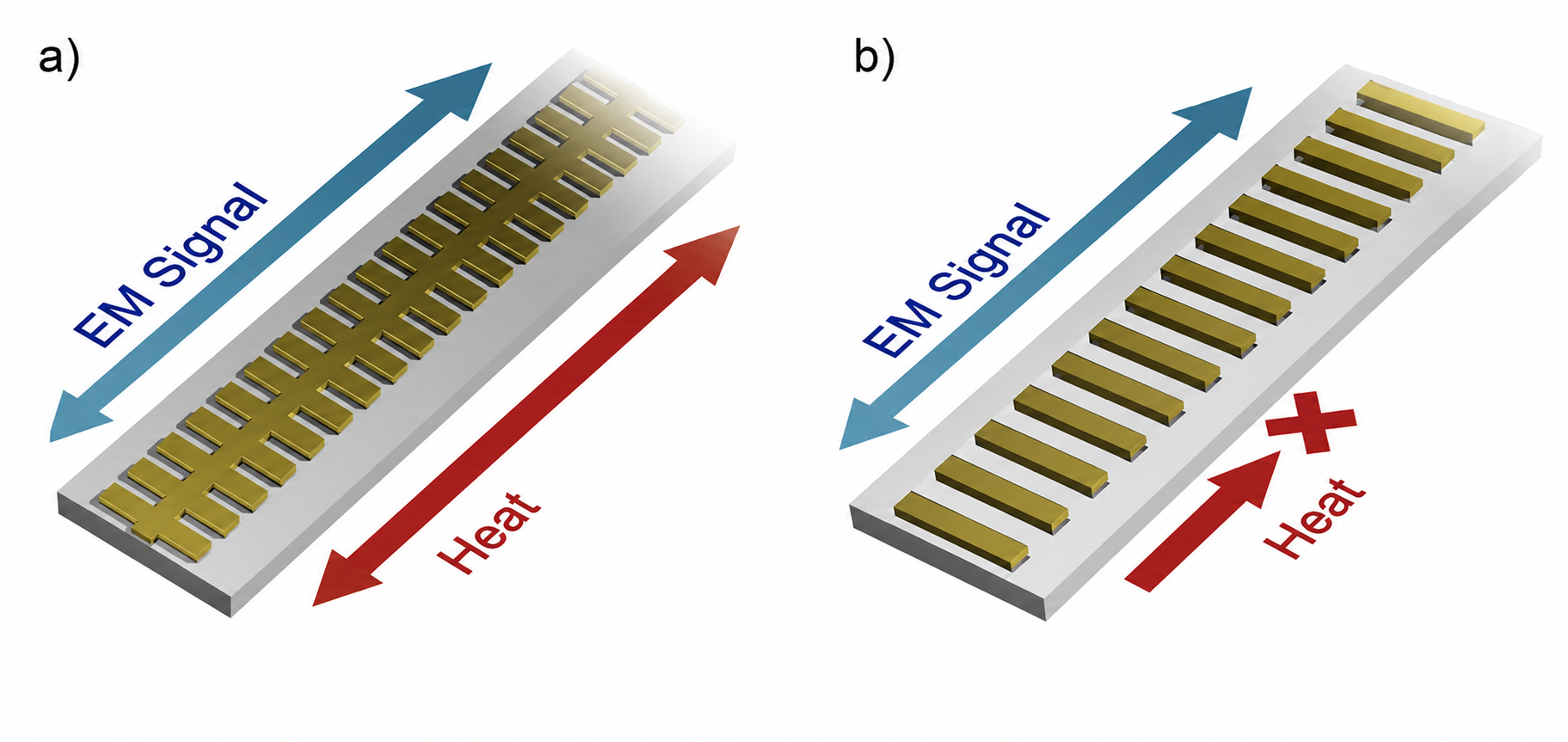}
\caption{(a) Conventional SSPP interconnect: electromagnetically and thermally conductive (b) Proposed SSPP interconnect: electromagnetically conductive, thermally opaque}
\label{fig:schematic}
\end{figure}

A single unpatterned metallic trace, however, cannot support a bound 
propagating mode at microwave frequencies. Surface plasmon polaritons (SPPs) 
are an example of a guided electromagnetic mode that propagates along a 
metal-dielectric interface with no return conductor, sustained by resonant 
plasma oscillations. Such a mode is inherently single-conductor in 
character. In practice, however, the plasma frequency of metals lies in the 
ultraviolet, confining SPPs to the hundreds-of-THz optical range and placing 
them far beyond microwave accessibility in uniform metallic structures. 
Pendry et al.~\cite{pendry2004} showed that subwavelength corrugations of 
depth $h$ imprinted on a metallic surface can spoof an analogous bound 
surface wave — a spoof surface plasmon polariton (SSPP) — with an effective 
plasma frequency $\omega_p = \pi c/2h$ tunable across the GHz-to-THz range 
entirely through geometry. This has motivated a wide variety of planar SSPP 
designs, including single- and double-edge corrugated metallic 
strips~\cite{shen2013, Erementchouk2016}, corrugated wires~\cite{maier2006}, 
staggered double-groove arrays~\cite{ye2016}, and complementary thin-film 
geometries~\cite{joy2018properties}, with dispersion and mode-converter 
characteristics studied in~\cite{zhang2018tap, unutmaz2022tap} and 
applications as feed structures and slow-wave radiators for antenna 
systems~\cite{wang2018tap, liu2021tap, zu2022tap}. These structures sustain 
a guided mode on a single patterned conductor with no ground plane. Yet the 
metallic structure in all reported SSPP designs remains a single continuous 
piece of metal from end to end (Fig.~\ref{fig:schematic}(a)): the 
corrugations shape the surface into grooves, but the underlying metal stays 
electrically joined at every point along the propagation direction. Thermal 
conductivity along the signal path therefore remains high. A notable 
exception is Gao~\textit{et al.}~\cite{gao2016chain}, who galvanically 
isolated the metallic elements themselves, arranging a chain of physically 
separate spoof plasmonic disk resonators that couple through weak near-field 
interaction rather than a continuous conductor. This near-field coupling is 
too weak to sustain useful transmission: the signal strength falls below 
$-10$~dB immediately after the first resonator and continues to drop with 
each additional resonator added to the chain, confining the design to a 
narrow bandwidth around each resonator's isolated characteristic frequency.

In this work, we show that high transmission efficiency and large fractional 
bandwidth can be achieved in a galvanically fragmented metallic trace by 
exciting a specific symmetry of the SSPP mode, in contrast to the weak 
near-field coupling and narrow bandwidth of prior isolated-resonator 
designs. When both lateral edges of the metallic trace are 
symmetrically corrugated (Fig.~\ref{fig:schematic}(a)), the opposing 
corrugation arrays couple and split the guided mode into even and 
odd-symmetric eigenmodes~\cite{Erementchouk2016, haus1991cmt}. The 
odd-symmetric eigenmode imposes a longitudinal electric field null along the 
central axis of the waveguide by antisymmetry. Consequently, the metallic 
bridges connecting adjacent unit cells along this null line can be removed 
entirely without perturbing the mode, reducing the corrugated strip to a 
chain of physically isolated metallic islands with no continuous conductor 
and no ground plane anywhere in the structure 
(Fig.~\ref{fig:schematic}(b)). Power is carried from island to island 
through resonant near-field coupling between adjacent corrugation cavities, 
with no conduction current flowing between them. The complete absence of any 
longitudinal conduction path suppresses thermal transport in the propagation 
direction. While the odd-symmetric eigenmode has been previously studied for 
differential signaling and mode multiplexing~\cite{oddmode2015, Du2020, 
He2025}, its implications for thermal isolation have not been recognized or 
exploited.

It is worth noting that other approaches to thermal leakage prevention in 
microwave interconnects have been explored, including replacing the metallic 
path with dielectric waveguides~\cite{galler2023, dey2018, 
lenschen2024dielectric} and introducing mechanical thermal breaks into 
rectangular hollow waveguides~\cite{melhuish2016high}. Dielectric waveguides 
suffer from weak modal confinement below 100~GHz, demanding prohibitively 
large cross-sections at lower microwave frequencies, while mechanical breaks 
are bulky three-dimensional structures that require precision alignment and 
cannot be cascaded without compounding insertion loss. The proposed 
isolated-island SSPP structure is a planar transmission line that supports 
low-frequency microwave propagation readily, with no mechanical assembly 
required. We demonstrate a 0.0031dB/mm insertion loss in a de-embedded waveguide, 12\% fractional bandwidth at 
C-band (5-6.5 GHz), a 40$\times$ reduction in thermal conductivity relative to a 
microstrip line of identical footprint, and 1~Gbps signal integrity in 
pulsed operation.

The remainder of this paper is organized as follows. 
Section~\ref{sec:principle} analyzes even/odd mode splitting, dispersion, 
transmission, and signal integrity in the isolated-island waveguide. 
Section~\ref{sec:thermal} presents a coupled electro-thermal analysis and 
compares thermal performance against conventional microstrip lines; 
Table~\ref{tab:comparison} benchmarks performance metrics against competing 
thermal management approaches. Section~\ref{sec:experimental} details the 
experimental methodology, including mode converter design, S-parameter 
measurements, and de-embedding.


\section{Working Principle of the Proposed Device}
\label{sec:principle}
\subsection{Electromagnetic Modes}


\begin{figure}[ht!] 
\centering
\includegraphics[width=3.45in]{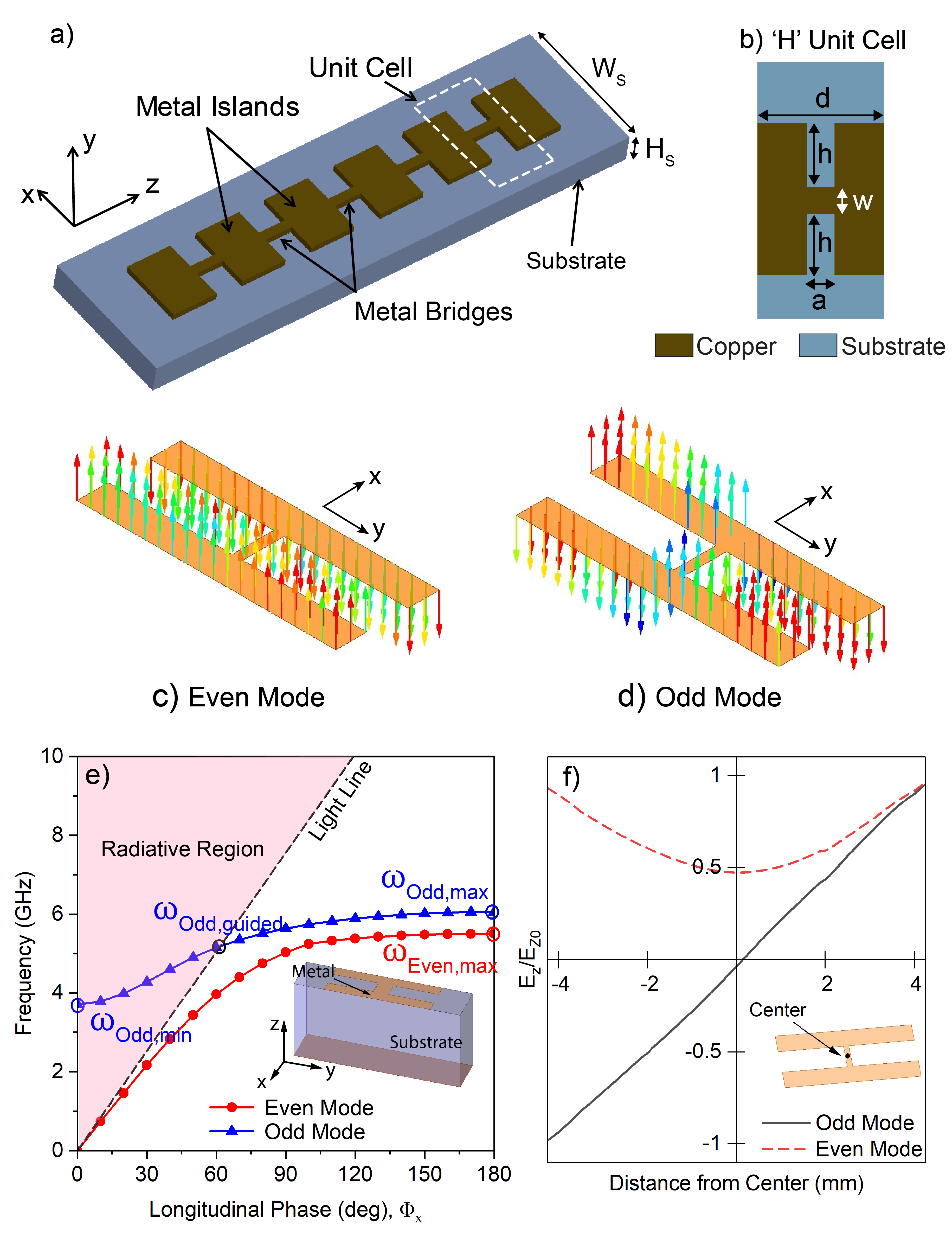}
\caption{(a) Schematic of SSPP interconnect with periodic array of metal islands and bridges (b) 'H' unit cell of bridged island SSPP waveguide with dimensions (c) direction of the electric field on the surface of the unit-cell of a bridged island for even and (d) odd mode (e) Dispersion diagram of even and odd mode, inset shows a unit cell (f) Variation of electric field component (E\textsubscript{z}) along the lateral axis (y-axis) of the unit cell.}
\label{fig:unitcellanalysis}
\end{figure}

\begin{figure}[ht!] 
\centering
\includegraphics[width=3.45in]{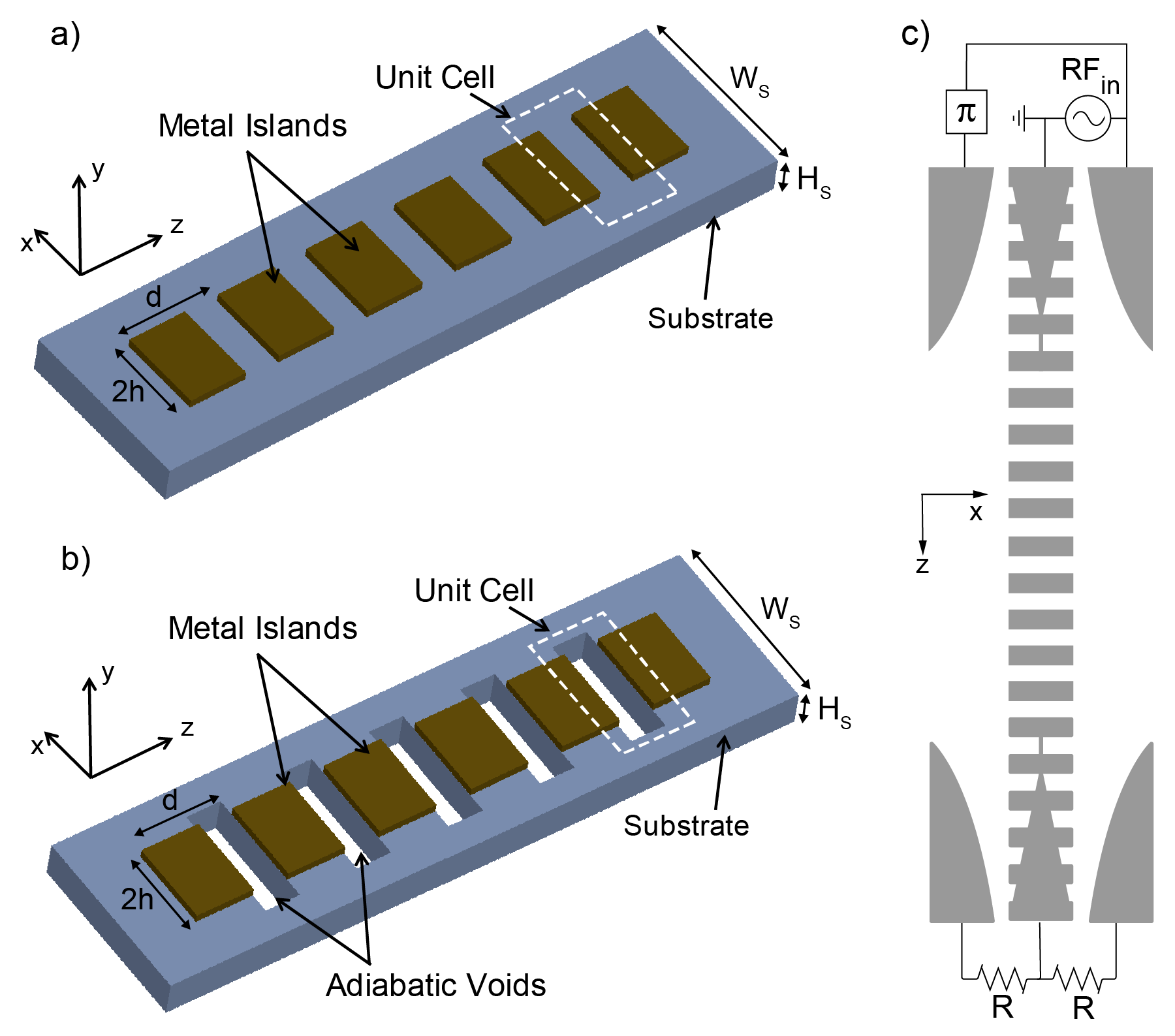}
\caption{Schematic of the two modified waveguides (a) Schematic of SSPP interconnect with periodic array of isolated metal islands (b) Schematic of SSPP interconnect with periodic array of isolated metal islands on voided substrate (c) Odd mode excitation with phase reversal of an incoming RF signal. }
\label{fig:twounitcells}
\end{figure}

The proposed periodic waveguide with tunable band and thermal insulation, shown in Fig.~\ref{fig:schematic}(b), carries the electromagnetic signal in the form of an odd-symmetric SSPP mode. We first briefly review the fundamentals of SSPPs before discussing mode symmetry. The name reflects its mimicry of the true `surface plasmon polariton' observed in noble metals near their plasma resonance ($\sim$200~THz), but realized at a designer-chosen frequency, typically GHz to sub-THz. A common realization is a metallic strip with both edges periodically corrugated (Fig.~\ref{fig:unitcellanalysis}(a)-(b)), where the corrugation depth $h$ sets the resonance frequency of the quarter-wavelength cavity formed by each groove, $\omega_r = \pi c / 2h$. An array of such cavities couples to form a propagating EM mode whose evanescent tail decays exponentially away from the metal-dielectric interface on all lateral sides.

For a metal strip corrugated on one edge with depth $h$, width $a$, and period $d$, the dispersion relation of the fundamental TM mode is
\begin{equation}
    \kappa= \sqrt{\beta^2-k_0^2}=k_0 \dfrac{a}{d}\,\text{sinc}^2\!\left(\dfrac{a\beta}{2}\right)\tan(k_0 h)
\end{equation}
where $k_0=\omega/c$ is the free-space wave-vector and $\kappa$ is the transverse decay constant of the evanescent wave. As $k_0 h \to \pi/2$, $\kappa$ increases asymptotically, indicating strong field confinement on the metal surface.

When both edges are symmetrically corrugated, the left- and right-edge corrugation arrays interact electromagnetically and split the eigenmode into an even mode (evanescent fields interfering in-phase) and an odd mode (interfering out-of-phase). The degree of splitting depends on the interaction strength between the two arrays, itself a function of frequency and the physical spacing between edges: for a waveguide width $(w+2h)$ much larger than $2h$, the even and odd modes are nearly degenerate. Viewing the structure as an array of metal islands connected by a central metallic bridge, narrowing the bridge width $w$ increases the mode-splitting. This is evident in the dispersion diagram of the unit cell (Fig.~\ref{fig:unitcellanalysis}(e)): the even-mode curve (red) stays below the light line from DC to the edge of the first Brillouin zone, so it can carry baseband data, though field confinement weakens at low frequency, raising the risk of signal perturbation from adjacent electronics. The odd-mode curve (blue) lies partly above and partly below the light line, crossing it at $\omega_{odd,guided}$ — a point tunable via the bridge width $w$ — making the odd mode guided from $\omega_{odd,guided}$ to the Brillouin-zone edge frequency $\omega_{odd,max}$.

\begin{figure}[ht!]
\centering
\includegraphics[width=3.45in]{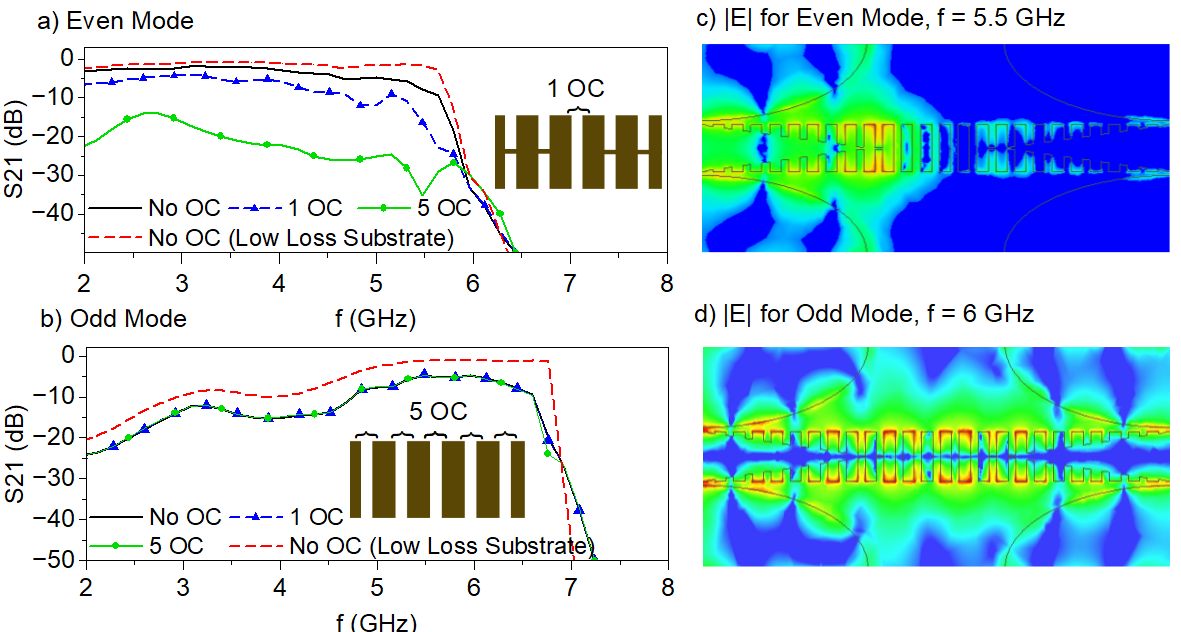}
\caption{Scattering parameter $S_{21}$ of (a) even mode (b) odd-mode. $S_{21}$ of several case of the waveguide-variants is analyzed: case when all metal islands are bridged, vs when some of the bridge is eliminated (c) Magnitude of electric field for even mode at resonance ($f = 5.5$ GHz) with 5 OC conditions (d) Magnitude of electric field for odd mode at resonance ($f = 6$ GHz) with 5 OC conditions.}
\label{fig:Transmission_with_OC}
\end{figure}

\begin{figure*}
\centering
\includegraphics[width=6.4in]{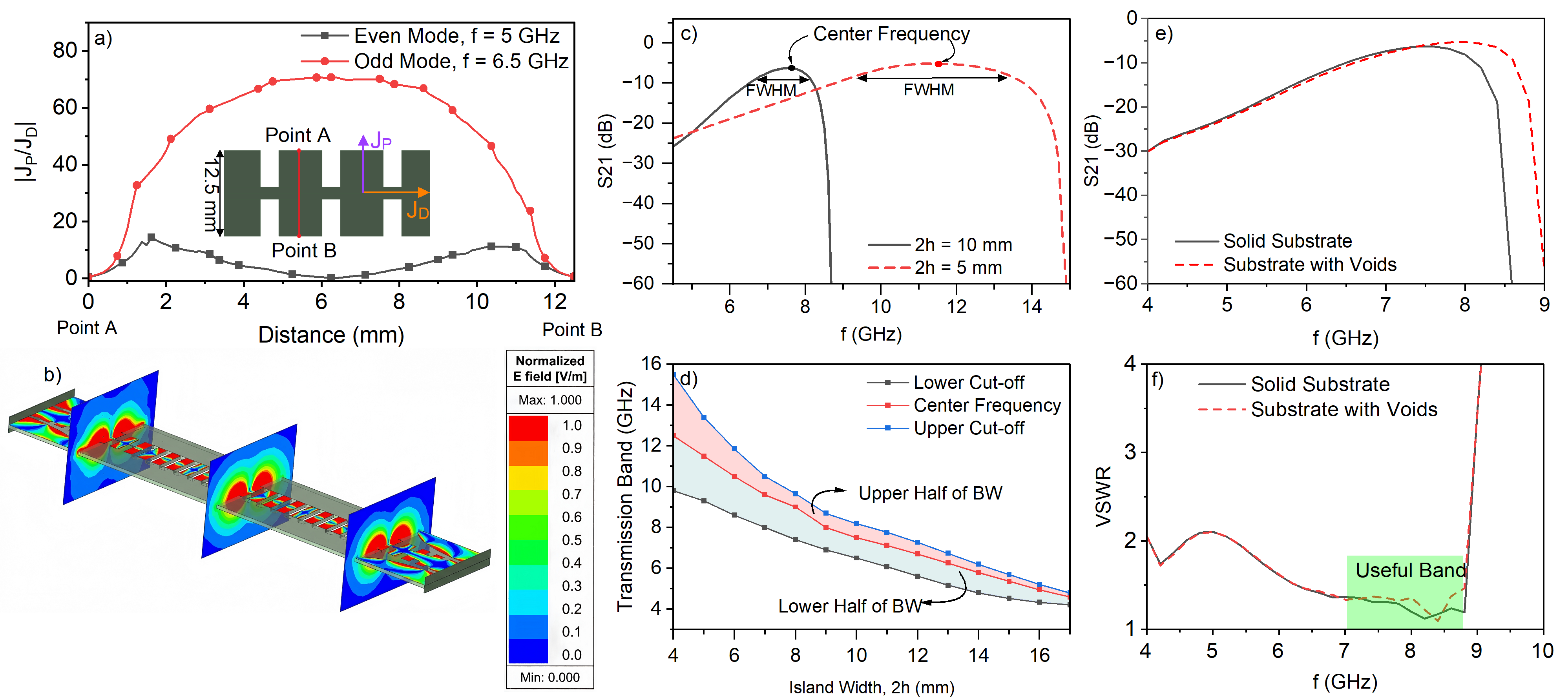}
\caption{(a) Ratio of polarization current to drift current for even and odd modes at resonance along the width of a 12.5 mm pitch waveguide (b) Cross section electric fields showing odd mode propagation through isolated islands (c) S-parameters of odd mode for two different island widths, with their center frequency and FWHM bandwidth (d) spectral tuning of the waveguide (e) S-parameters for a waveguide for solid and void-etched substrate (f) Voltage standing wave ratio (VSWR) for the two types of substrates.}
\label{fig:islandheighteffect}
\end{figure*}

The guiding property of the even and odd-symmetric modes becomes apparent in the scattering parameters. Fig.~\ref{fig:twounitcells}(c) shows the excitation mechanism: the periodic SSPP waveguide is flanked at both terminals by a vivaldi-antenna-like converter used for mode-matching. The RF source's live terminal feeds one wing of the converter, with its ground connected to the waveguide center; the other wing connects to the live terminal via a delay line providing a $2\pi$ phase shift for even-mode excitation and $\pi$ for odd-mode excitation. Fig.~\ref{fig:Transmission_with_OC}(a)-(b) shows the resulting $S_{21}$ for each mode. As predicted from the dispersion analysis, the even mode transmits well ($|S_{21}|>-5$~dB) from low frequency up to its cut-off at 5.75~GHz, while the odd mode has a narrower band with a higher cut-off at 6.5~GHz, transmitting well only within 5--6.5~GHz.

Although the odd-symmetric mode has a narrower bandwidth and is harder to 
excite in pure form (discussed further in the experimental section), it has 
a unique feature exploitable for thermally resistive interconnects: the 
narrow metallic bridges between adjacent islands can be eliminated entirely 
without perturbing odd-mode propagation. Fig.~\ref{fig:unitcellanalysis}(c)-(d) 
plots the dominant electric field component ($E_z$) on the unit-cell surface 
for the even- and odd-symmetric modes; the odd-mode field flips sign across 
the waveguide width along $y$, vanishing at the centerline 
(Fig.~\ref{fig:unitcellanalysis}(f)). Removing the centerline bridge 
therefore leaves the odd mode unperturbed, reducing the corrugated strip to 
a periodic array of physically isolated metallic islands 
(Fig.~\ref{fig:twounitcells}(a)). Unlike capacitively-coupled discontinuous 
designs, where only the signal trace is interrupted while a continuous 
ground plane beneath the substrate still completes the return-current path, 
the proposed structure has no ground plane at all: every conductive path, 
forward and return, is broken. This concept of a 
spoof plasmonic waveguide formed by physically discrete, periodically 
arranged 3D metal slabs was first proposed in~\cite{Erementchouk2016} (where 
the PI of this work was a co-author) as an "open-groove" structure, though 
its realization on a planar geometry was not established at the time; we 
first demonstrated the odd-symmetric mode on a periodically fragmented 
planar strip in~\cite{jawad2025}. Fig.~\ref{fig:Transmission_with_OC} 
confirms this by comparing the effect of bridge removal on even- and 
odd-mode transmission.

For this analysis, we used a waveguide with 25 rectangular copper islands, period $d=5$~mm and $h=6$~mm; islands are $2h+w=12.5$~mm wide and $d-a=3$~mm long, with $a=2$~mm spacing between adjacent islands, on a 1.6~mm thick, 40~mm wide FR4 substrate. No ground plane is used, since such a periodically patterned metal wire supports a surface-bound mode and can propagate signal without one~\cite{joy2019}. While the waveguide geometry can be tailored to any arbitrary transmission band, our specific dimensions were set by the manufacturer's fabrication constraints and the measurement limits of the available VNA.

The design was numerically analyzed with bridges present and absent, the latter labeled Open Contact (OC) in Fig.~\ref{fig:Transmission_with_OC}, where $\#N$~OC denotes $N$ bridges removed, isolating the neighboring islands. While $|S_{21}|$ for the even mode drops sharply as more bridges are removed, the odd-mode $S_{21}$ remains completely unaltered. Fig.~\ref{fig:Transmission_with_OC}(c) shows the absence of electric field beyond the isolated bridges for the even mode, confirming it fails to propagate, whereas Fig.~\ref{fig:Transmission_with_OC}(d) shows odd-mode transmission persisting even with 5 bridges removed. The odd-mode $S_{21}$ reaches a maximum of $-4.6$~dB at 5.5~GHz on FR4; this could approach 0~dB with a lower-loss substrate.

This behavior is explained by the current distribution on the metal islands (Fig.~\ref{fig:islandheighteffect}(a)). Conventional interconnects rely on drift current along the propagation direction, so they cannot carry a signal across a broken wire. The SSPP interconnect instead relies on polarization current transverse to propagation, with longitudinal current nearly zero except at the edges, allowing signal transmission across an open-circuit junction. At resonance (6.5~GHz), the odd-mode transverse current exceeds the longitudinal current by more than two orders of magnitude (Fig.~\ref{fig:islandheighteffect}(a)), and the cross-sectional field distribution (Fig.~\ref{fig:islandheighteffect}(b)) confirms the guided field vanishes at the waveguide center, making it insensitive to removal of galvanic connections between islands.

Fig.~\ref{fig:islandheighteffect}(c) shows the odd-mode S-parameter for two lateral sizes, $2h=10$~mm and $2h=5$~mm: scaling down the lateral size raises the spoof plasmonic resonance and shifts the transmission window higher in frequency. Characterizing this window by its central frequency and $-3$~dB upper/lower cut-offs, Fig.~\ref{fig:islandheighteffect}(d) shows how all three scale with $2h$, demonstrating spectral tunability via geometry — useful for routing a microwave control signal to low-temperature electronics while suppressing out-of-band noise.

To further reduce heat propagation, we partially etched the substrate between islands to form adiabatic voids of $2\times10$~mm (Fig.~\ref{fig:twounitcells}(b)), retaining the substrate along the lateral edges as a mechanical support rail. To limit the electromagnetic reflection that abrupt voids can introduce via mode mismatch, the substrate was kept thin ($1$~mm). Fig.~\ref{fig:islandheighteffect}(e)-(f) compares $S_{21}$ and VSWR for solid versus voided substrates: since the substrate is much thinner than the operating wavelength, the voids cause only a negligible rise in reflection, confirming good EM signal integrity.

\subsection{Waveguide Bend Analysis}

\begin{figure*}
\centering
\includegraphics[width=6.4in]{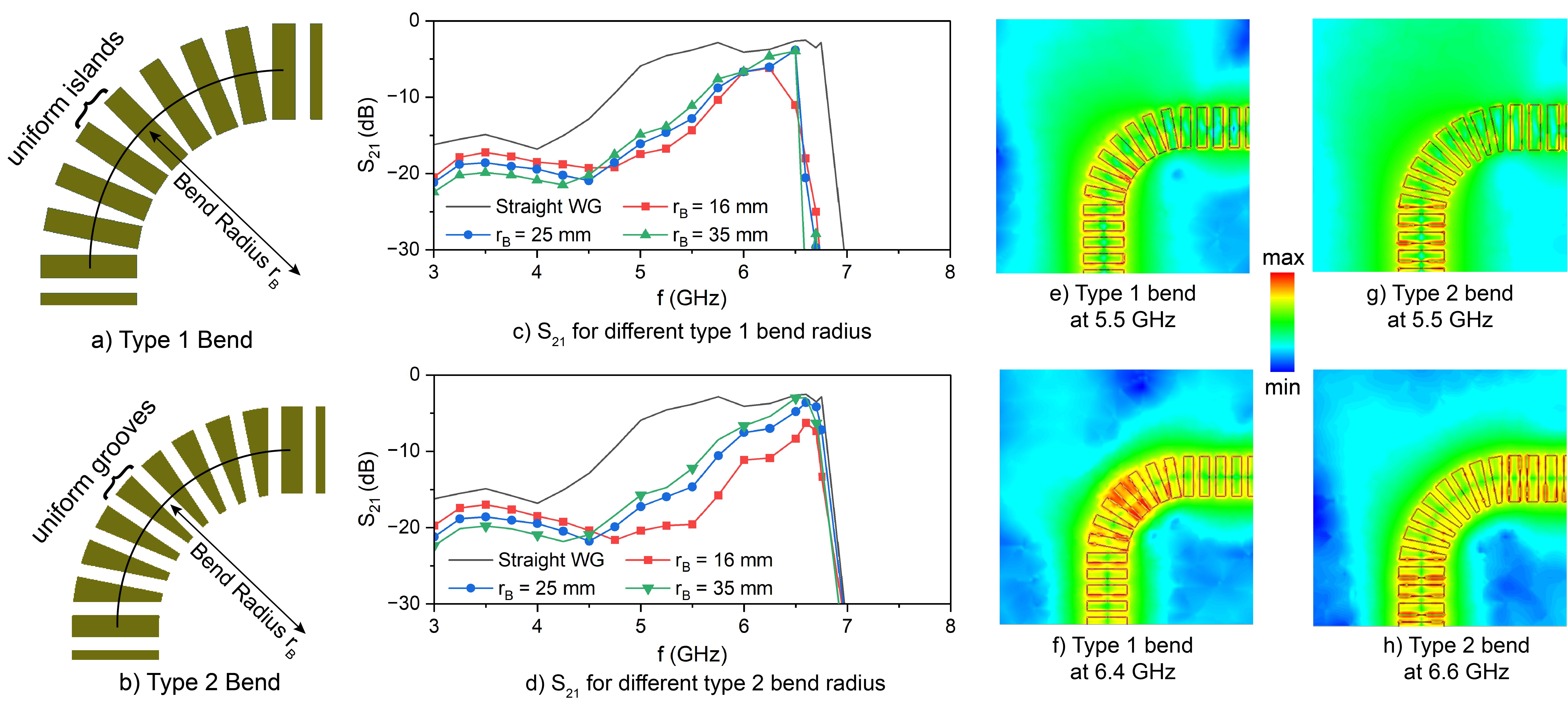}
\caption{90\textsuperscript{o} bends in proposed waveguide (a) Type 1 bend: islands in the bend are uniform in the size, grooves fan out. (b) Type 2 bend: grooves in the bend are uniform, metallic islands fan out. (c) S\textsubscript{21} of the 90\textsuperscript{o} bends for different bend radii for type 1 bends and (d) type 2 bends. (e)-(h) Logarithm of total near field power density for two types of bends near upper and lower cut-off frequencies of the waveguide. }
\label{fig:waveguidebends}
\end{figure*}

To assess the robustness of the proposed waveguide for practical routing, we analyze its electromagnetic performance under curved trajectories. We evaluate two $90^{\circ}$ bend geometries, shown in Fig.~\ref{fig:waveguidebends}(a)-(b). In Type 1, the metallic islands keep a constant rectangular footprint and are simply rotated along the bend radius $r_B$, leaving non-uniform, outward-widening grooves. In Type 2, the islands are tapered to maintain a constant groove width $a$ throughout the curvature, resulting in a pie-shaped geometry. Both types are evaluated at bend radii of 16, 25, and 35~mm and compared against a straight reference waveguide.

Sharp curvature introduces impedance mismatch at the bend input, and transmission magnitude declines as the bend radius decreases for both types. While the straight waveguide supports a 5--6.75~GHz passband, curvature constricts the effective bandwidth, primarily through an upshift in the lower cut-off frequency: at lower frequencies the longer wavelengths are insufficiently confined to follow the curvature and radiate tangentially into the surrounding medium, visible in the near-field power density at 5.5~GHz (Fig.~\ref{fig:waveguidebends}(e),(g)). Near the spoof plasmon resonance, field confinement is stronger and the signal navigates the $90^{\circ}$ turn with minimal degradation, as seen for the 25~mm bends in Fig.~\ref{fig:waveguidebends}(f),(h). For this radius, the $-3$~dB bandwidth reduces to approximately 0.7~GHz for Type 1 and 0.85~GHz for Type 2.

Since the spoof plasmon cut-off frequency is set by groove geometry, Type 2 bends, which preserve a constant groove width throughout the curvature, maintain a stable upper cut-off frequency (Fig.~\ref{fig:waveguidebends}(h)). Type 1 bends, in contrast, develop wider grooves on the outer edge of the curve, which shift the cut-off downward. Despite these radiation losses, both bend geometries retain a functional bandwidth close to that of the straight waveguide.

\subsection{Time Domain Signal Analysis}
A common concern with metamaterial-based interconnects is dispersion: different frequency components travel at different speeds, distorting pulse shape. To assess signal integrity, we analyzed group delay, $\tau_g=-d\theta/d\omega$, where $\theta$ is the unwrapped phase of $S_{21}$, related to the longitudinal wave-vector as $\theta=\beta L$. Since the SSPP waveguide confines the field more strongly as frequency approaches the spoof plasmonic resonance ($\beta$ increases as $\omega\rightarrow \pi c/2h$), group delay is expected to rise with frequency.

\begin{figure}
\centering
\includegraphics[width=3.45in]{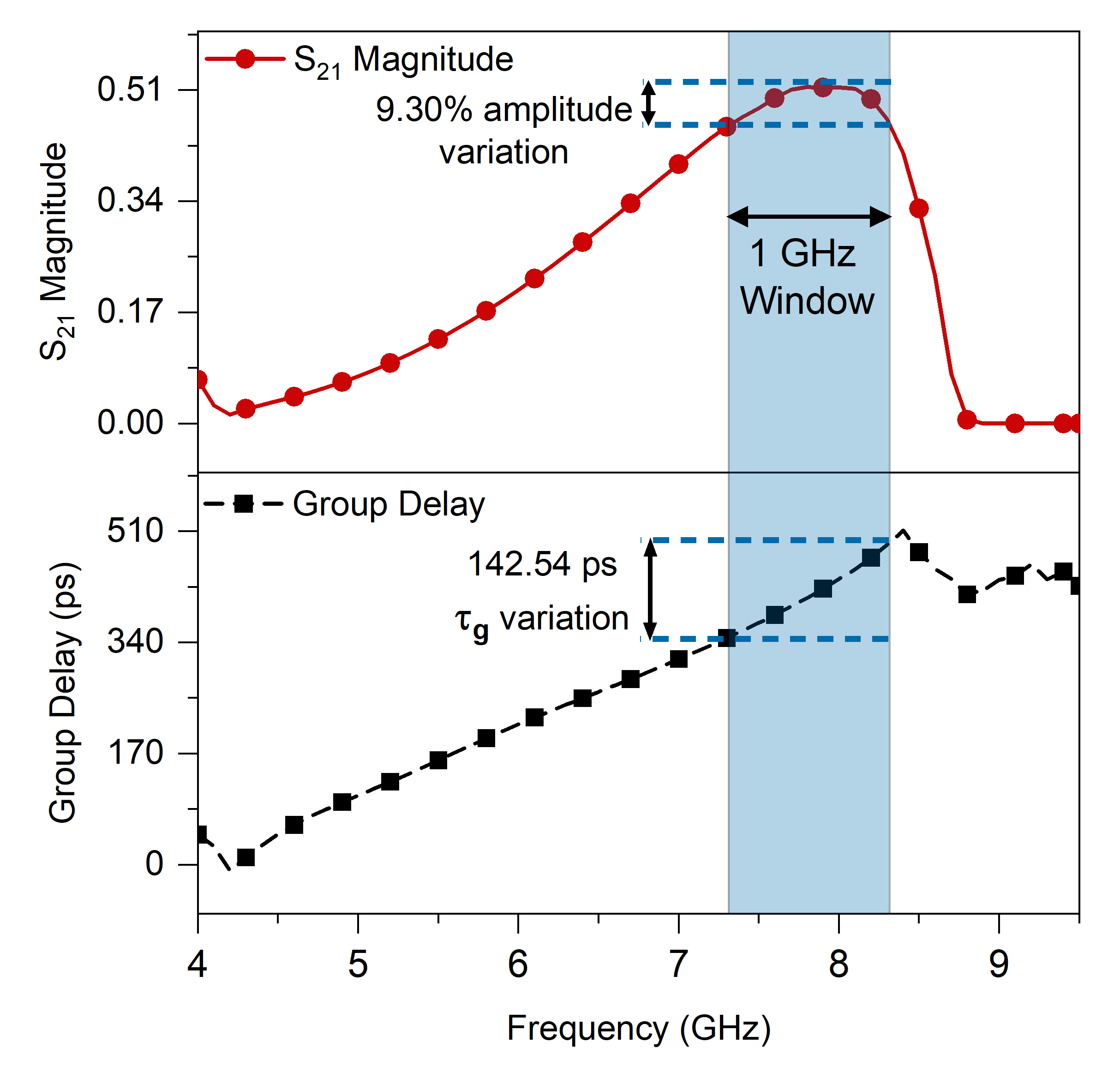}
\caption{Magnitude of Transmission coefficient $|S_{21}|$ and the corresponding group delay $\tau_g$ of odd-symmetric SSPP mode for waveguide width of 10 mm.}
\label{fig:group_velocity}
\end{figure}

Fig.~\ref{fig:group_velocity} shows $|S_{21}|$ and $\tau_g$ for the odd mode of an isolated-island SSPP waveguide with 10~mm island width. Within a 1~GHz window centered at 7.9~GHz, $|S_{21}|$ varies by approximately 10\% and $\tau_g$ by $\Delta\tau_g\approx140$~ps — small relative to the 1~ns pulse width of a 1~Gbps signal, indicating good signal integrity for digital transmission at or below 1~Gbps via amplitude modulation at this carrier frequency. Shaping the binary pulse with a root-raised-cosine (RRC) filter can further improve integrity by limiting its spectral span.

\begin{figure}
\centering
\includegraphics[width=3.45in]{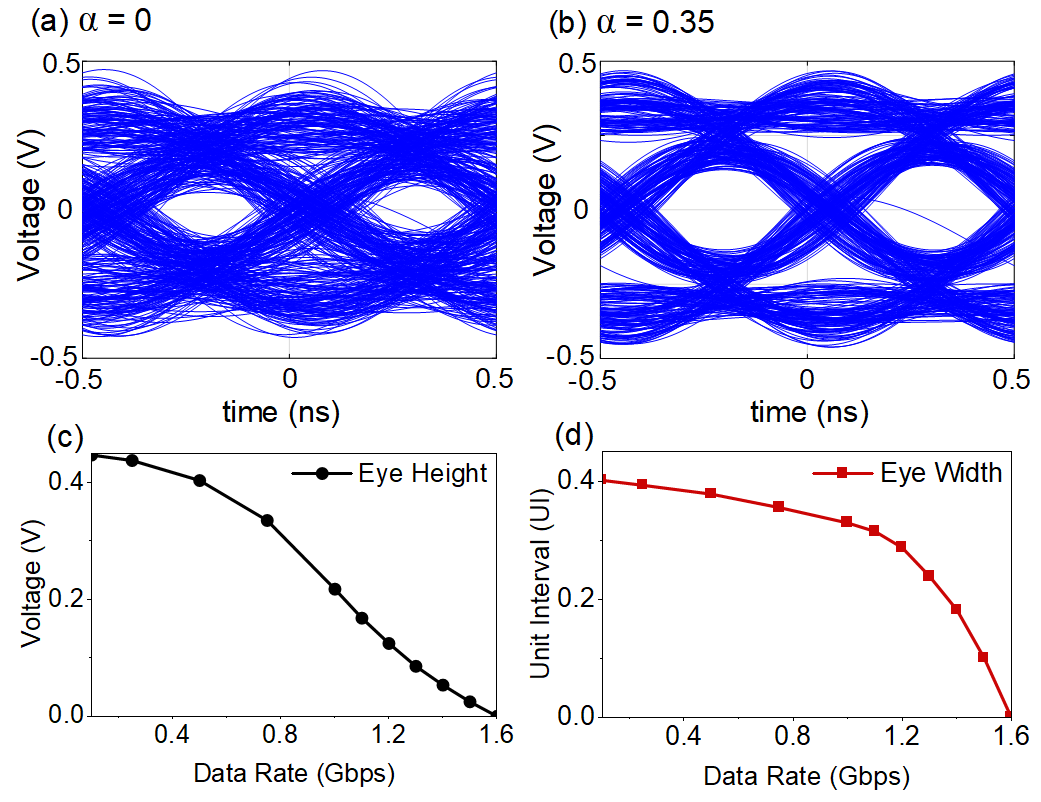}
\caption{Eye-diagram of SSPP output signal after AM demodulation. The input signal is random binary data of 1 Gbps, AM modulated by a carrier of 7.9 GHz and shaped by RRC filter with roll-off parameter (a) $\alpha = 0$ and (b) $\alpha=0.35$ (c) Eye height and (d) eye width for different data rates.}
\label{fig:eye-diagram}
\end{figure}

For a 1~Gbps random binary bitstream, RRC-shaped and AM-modulated on a 7.9~GHz carrier, Fig.~\ref{fig:eye-diagram} compares the demodulated eye-diagram for two roll-off factors. With $\alpha=0$ (sharp cut-off, Fig.~\ref{fig:eye-diagram}(a)), the eye-height is 0.1~V and eye-width 0.2~UI; the sharp cut-off broadens the pulse in time and introduces inter-symbol interference. Increasing the roll-off to $\alpha=0.35$ (Fig.~\ref{fig:eye-diagram}(b)) noticeably improves both eye-height and eye-width, confirming good signal integrity potential. Fig.~\ref{fig:eye-diagram}(c)-(d) shows eye-height and eye-width declining as data rate increases from a few Mbps to a few Gbps at the 7.9~GHz carrier, with the eye fully closing beyond 1.6~Gbps.

\section{Coupled Electro-Thermal Analysis}
\label{sec:thermal}
\begin{figure*}
\centering
\includegraphics[width=6.9in]{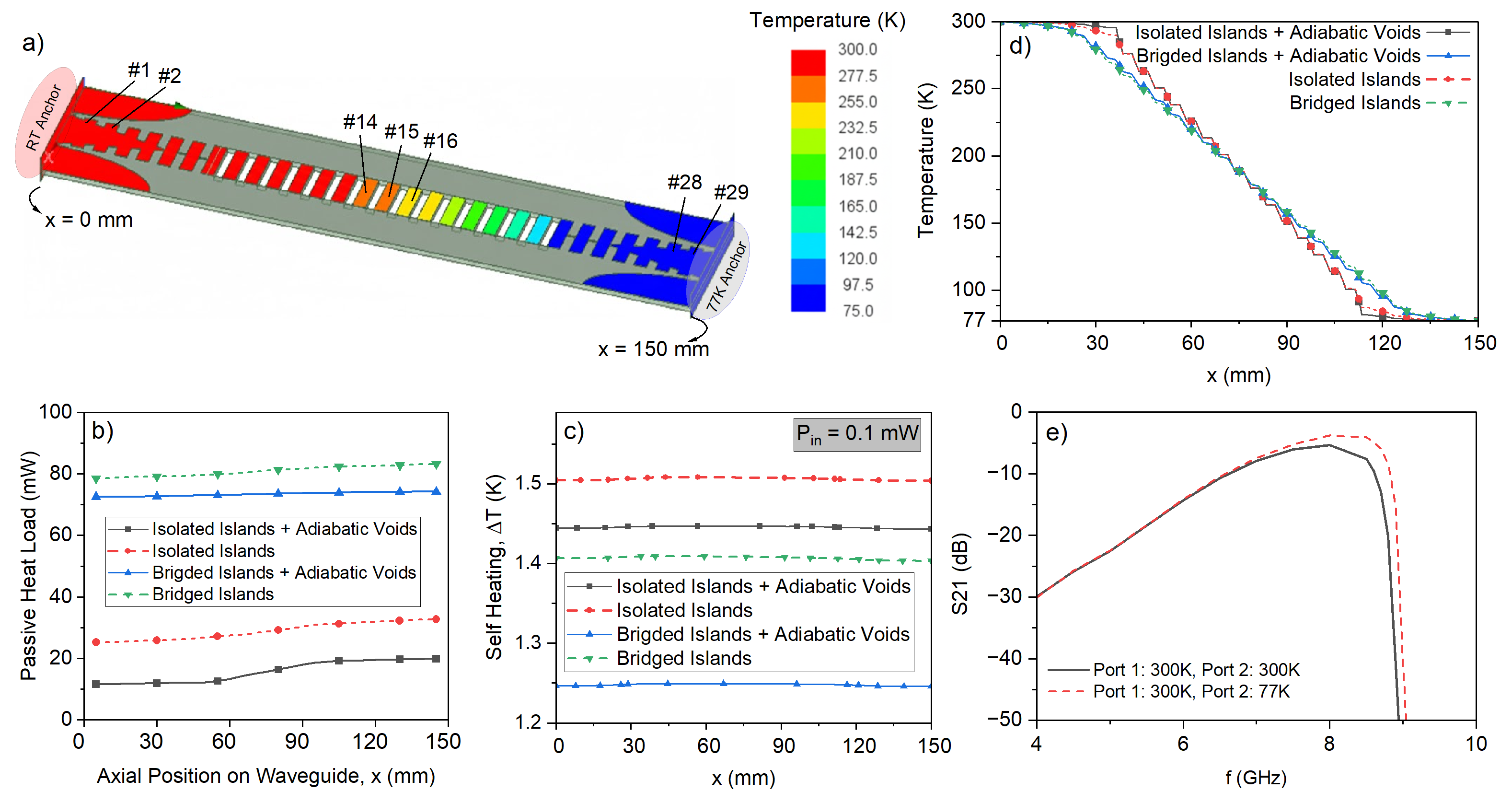}
\caption{(a) Temperature Profile of a SSPP interconnect with isolated islands with adiabatic voids (b) Calculated passive heat load of islands, expressed in milliwatts (c) Calculated temperature increase due to self heating for 0.1 mW active load (d) Temperature along the longitudinal centerline of the interconnect for the four interconnect variants (e) Simulated comparison between S-parameters under room temperature (300K), vs under an applied thermal gradient of 300K-77K.}
\label{fig:islanddetails}
\end{figure*}

We performed coupled electro-thermal multiphysics simulation to evaluate (1) how the waveguide's thermal conductivity changes with the geometry of the electromagnetic structure, and (2) how the resulting temperature distribution feeds back into its electromagnetic transport. Transient 3D thermal simulations were conducted in Ansys Icepak, performing a conjugate heat transfer (CHT) analysis by solving the Navier-Stokes and energy equations simultaneously. To single-out the contribution of conductive heat transport along the waveguide, the surrounding medium was modeled with negligible thermal conductivity, density, and specific heat ($10^{-10}$~W/m$\cdot$K, $10^{-10}$~kg/m$^3$, $10^{-10}$~J/kg$\cdot$K, respectively), suppressing convective exchange with the environment. To evaluate thermal-isolation performance under a representative large temperature differential, we simulated a 150~mm interconnect spanning a 300~K to 77~K gradient — a wide swing chosen to stress-test the structure's thermal decoupling rather than to target any specific operating environment. Copper was modeled with $\kappa=400$~W/m$\cdot$K, $\rho=8933$~kg/m$^3$, $C_p=385$~J/kg$\cdot$K, and the substrate with $\kappa=0.3$~W/m$\cdot$K, $\rho=1900$~kg/m$^3$, $C_p=1150$~J/kg$\cdot$K; all properties are isotropic.

Islands are indexed $\#1$--$\#29$ from the hot end (300~K) to the cold end (77~K) (Fig.~\ref{fig:islanddetails}(a)), with islands $\#7$--$\#23$ physically isolated and the remainder electrically connected to form the adapter. At steady state, the thermal conductivity $k$ of the islands is calculated as
$k = \frac{Q \Delta x}{A(T_2 - T_1)},$
where $Q$ is the steady-state heat flux, $\Delta x = 150$~mm is the analyzed segment length, $A = 10$~mm$^2$ is the cross-sectional area (10~mm island width $\times$ 0.1~mm Cu thickness, plus 1~mm substrate thickness), and $T_2$, $T_1$ are the hot- and cold-plate temperatures.

To isolate the contributions of metal-island fragmentation and substrate voiding to heat conduction, we performed thermal analysis on four waveguide variants: variant-1, isolated islands with periodic substrate voids (Fig.~\ref{fig:twounitcells}(b)), the primary proposed design; variant-2, isolated islands on a solid substrate (Fig.~\ref{fig:twounitcells}(a)); variant-3, electrically bridged islands on a solid substrate (Fig.~\ref{fig:unitcellanalysis}(a)); and variant-4, bridged islands with periodic substrate voids. Comparing variant-1 to variant-2 isolates the contribution of the substrate alone; variant-2 to variant-3 isolates the contribution of the metallic bridge; and variant-3 to variant-4 reveals whether substrate voiding remains effective when a dominant parallel thermal path (the bridge) is present. For reference against conventional interconnects, we also analyzed a continuous, uncorrugated metallic strip of identical footprint on a solid substrate.

\begin{figure}
\centering
\includegraphics[width=3.45in]{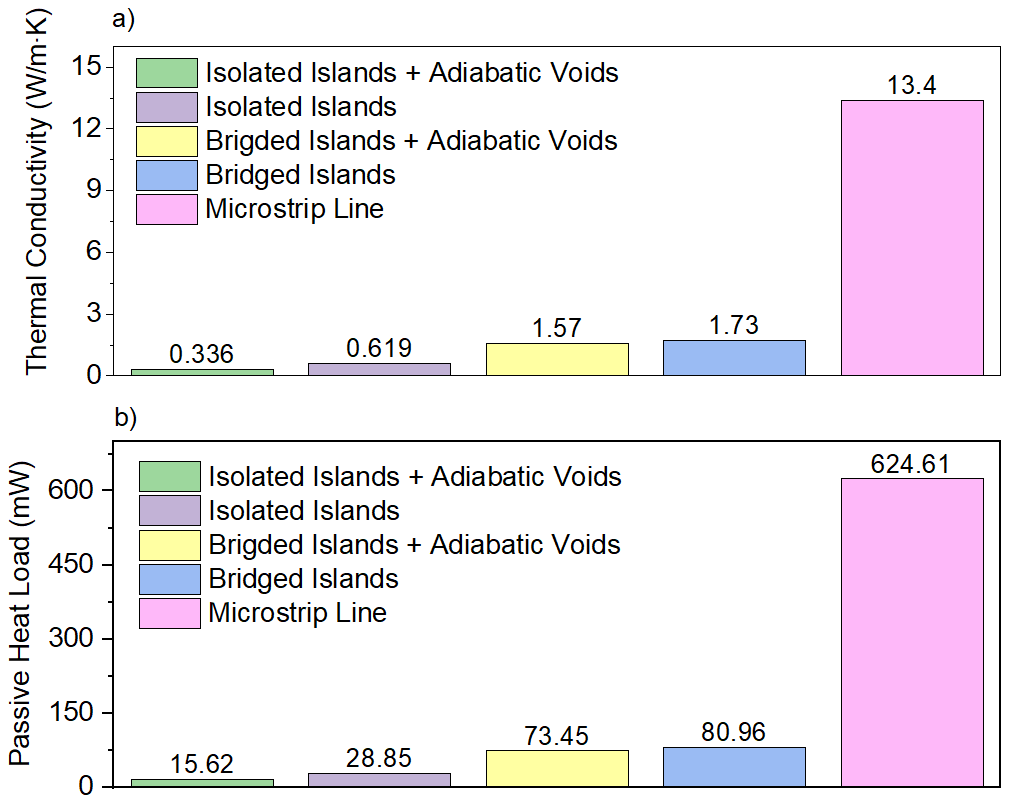}
\caption{(a) Simulated thermal conductivity and (b) calculated passive heat load of all five analyzed interconnects, evaluated across a 300K to 77K applied gradient.}
\label{fig:thermalsummary}
\end{figure}

Fig.~\ref{fig:islanddetails}(b) shows that the passive heat load is highest for variant-3 (80~mW), as the metallic bridges form a direct conduction path between islands. Introducing substrate voids beneath the bridges (variant-4) reduces this to 72~mW. Removing the bridges entirely (variant-2) drops the heat load by more than a factor of 3, and combining island isolation with substrate voids (variant-1) brings it to a minimum of 10~mW, an 8$\times$ reduction relative to the fully bridged case. Fig.~\ref{fig:thermalsummary}(a)-(b) summarizes the effective thermal conductivity and passive heat load across all variants: the proposed island-SSPP with substrate voids reduces thermal conductivity by a factor of 40 relative to a continuous metal strip of the same footprint.

Isolating the metallic islands introduces one minor thermal trade-off: generated heat within an island cannot escape as readily as in a continuous conductor, leading to self-heating during EM signal flow. This effect is appreciable mainly at higher injected power; for low-power signals ($<1$~mW), self-heating remains small relative to the thermal leakage suppression achieved by island isolation. The continuous dielectric rail along both lateral sides of the waveguide, included for mechanical support, also aids heat spreading and further limits self-heat trapping. Fig.~\ref{fig:islanddetails}(c) shows that, for a 0.1~mW EM signal, the steady-state self-heating of isolated islands raises the temperature only 1.5~K above baseline, barely 0.3~K above that of electrically bridged islands.

Since the EM simulations in Section~\ref{sec:principle} assumed isothermal conditions, we refined the analysis by incorporating the temperature gradient along the waveguide (Fig.~\ref{fig:islanddetails}(d)), which affects the thermal and electrical properties of the metal and substrate~\cite{siu1976}. The resulting electro-thermal coupled analysis shows only a nominal ($\sim 0.2$~GHz) shift in the SSPP cut-off frequency (Fig.~\ref{fig:islanddetails}(e)). Repeating this analysis for a different temperature gradient (150~K to 50~K) yields an effective thermal conductivity nearly independent of the exact gradient, as expected.

\begin{table*}[t]
\caption{COMPARISON OF GUIDED-WAVE INTERCONNECT TECHNOLOGIES FOR THERMALLY CONSTRAINED SIGNAL ROUTING}
\label{tab:comparison}
\centering
\small
\begin{tabular}{|l|c|c|c|c|c|c|}
\hline
\textbf{Technology} & \textbf{\begin{tabular}[c]{@{}c@{}}Frequency\\ Band\end{tabular}} & \textbf{\begin{tabular}[c]{@{}c@{}}Insertion\\ Loss\end{tabular}} & \textbf{\begin{tabular}[c]{@{}c@{}}Passive Heat \\Load (mW)\end{tabular}} & \textbf{\begin{tabular}[c]{@{}c@{}}Th. Conductivity \\(W/m$\cdot$K)\end{tabular}} & \textbf{\begin{tabular}[c]{@{}c@{}}Conversion Power\\ Overhead\end{tabular}} & \textbf{\begin{tabular}[c]{@{}c@{}}Primary\\ Limitation\end{tabular}} \\ \hline
\textbf{Dielectric WG} \cite{lenschen2024dielectric} & $>80$ GHz & $\mathbf{0.6}$ \textbf{dB/m} & $\mathbf{1.4}$ $\mu$\textbf{W}$^{\mathrm{a}}$ & $3$ & $>140$ mW~\cite{sarkas2009w}$^{\mathrm{b}}$ & Requires Upconversion \\ \hline
\textbf{Mech. Break} \cite{melhuish2016high} & Ka / Q Band & $0.1$ dB/break & $501$ mW $^{\mathrm{c}}$ & $0.2$ $^{\mathrm{d}}$ & --- $^{\mathrm{e}}$ & Bulky 3D Alignment \\ \hline
\textbf{This Work (SSPP)} & \textbf{C / X-Band} & $\sim \mathbf{3.1}$ \textbf{dB/m}$^{\mathrm{f}}$ & $\mathbf{15.62}$ \textbf{mW} & $0.336$ & \textbf{None}$^{\mathrm{g}}$ & \textbf{Converter Loss} \\ \hline
\end{tabular}
\footnotesize{\\ $^{\mathrm{a}}$293K-45K gradient. $^{\mathrm{b}}$ At room temperature. $^{\mathrm{c}}$$300K-50K$ gradient per thermal break. $^{\mathrm{d}}$Recalculated from $65K-2.7$ K temperature gradient data. $^{\mathrm{e}}$Comparable converter/transition power overhead not independently reported for this technology. $^{\mathrm{f}}$Based on simulated de-embedded data. $^{\mathrm{g}}$ Provided the operating frequencies are comparable to digital clock frequency.

}
\end{table*}

While the proposed discontinuous SSPP transmission line topology can be geometrically tailored to support other frequency bands, we anticipate its advantages to be most pronounced in the lower microwave spectrum (C/X-band). In this regime, metallic interconnects show low ohmic loss, whereas competing dielectric waveguides suffer from weak mode confinement, necessitating prohibitively large waveguide cross-sections to prevent radiative loss. In contrast, the proposed SSPP structure maintains strong confinement and compact planar dimensions comparable to standard metallic transmission lines, while offering thermal isolation previously attainable only with dielectrics. By aligning the passband with the desired signal band through geometric scaling of the corrugation depth, the interconnect inherently rejects out-of-band signal content as part of its guided-mode response, without requiring a separate filtering stage or the power penalty associated with upconverting to mm-wave frequencies. Its discrete-island structure also drastically lowers the thermal conductivity compared to standard microstrip lines. Thus the proposed design introduces a guided-wave architecture that combines the planar integrability of metallic transmission lines with the thermal resistance previously achievable only in dielectric structures, albeit at the cost of additional insertion loss relative to continuous metallic lines. Table~\ref{tab:comparison} compares the proposed periodically discontinuous SSPP transmission line against other thermally resistive microwave interconnect approaches.

\section{Experimental Verification}
\label{sec:experimental}

\begin{figure*}
\centering
\includegraphics[width=6.8in]{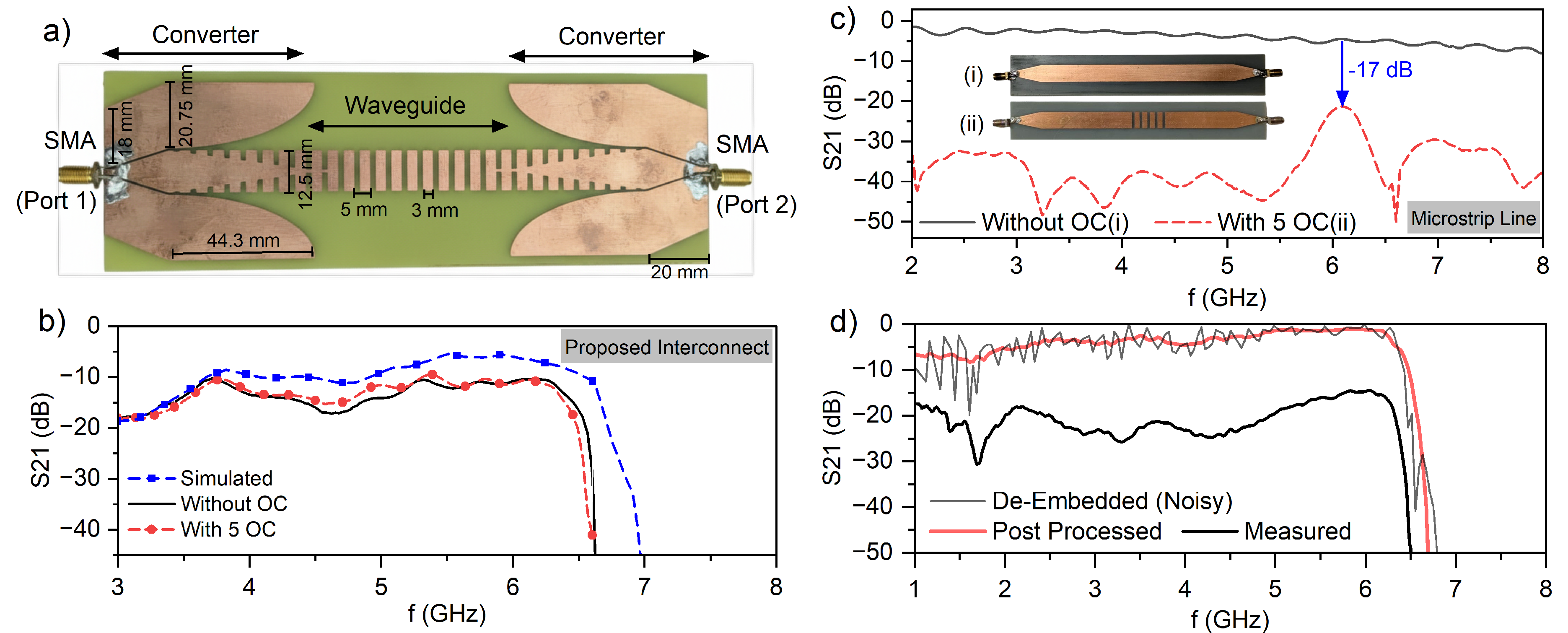}
\caption{(a) Analyzed sample of isolated island waveguide with 10 islands and converters to generate pure odd mode (b) Transmission property of fabricated isolated island waveguide and bridged island waveguide, when only odd mode is excited (c) Effect of conductor discontinuity in a microstrip line (d) De-Embedded S-parameter from experiment for a isolated island SSPP interconnect on solid substrate.}
\label{fig:experimental}
\end{figure*}

A key measurement challenge is converting the TEM mode of the VNA coaxial 
feed into the odd-symmetric SSPP mode. Since the odd-symmetric mode requires 
anti-phase excitation at the two lateral edges of the waveguide, the 
converter shown in Fig.~\ref{fig:experimental}(a) uses electromagnetic 
induction to bifurcate an incoming TEM wave into two voltage sources 
naturally $180^\circ$ out of phase, selectively coupling power into the odd 
mode while suppressing the even mode.

To verify propagation through physically isolated islands, we fabricated two 
structures on a 1.6~mm thick FR4 substrate: a 165~mm, 25-unit-cell bridged 
island waveguide (no open-contact (OC) conditions), and an otherwise 
identical waveguide with 5 isolated islands (5 OC conditions), created by 
removing bridges between neighboring unit cells. S-parameters were measured 
with a Copper Mountain VNA. Fig.~\ref{fig:experimental}(b) shows the 
measured $|S_{21}|$ for both configurations; the no-OC data agree well with 
simulation, aside from additional loss from mode conversion and SMA 
connectors. As predicted, the transmission response is nearly identical 
regardless of whether the bridges are present or removed.

For comparison, we fabricated two microstrip lines of similar pitch, 
introducing identical discontinuities in one. As shown in 
Fig.~\ref{fig:experimental}(c), this lowers transmission by at least 17~dB 
across the measured spectrum, confirming that our waveguide propagates signal 
through discontinuities that a standard microstrip line cannot.

The measured data remained lossy due to inefficient TEM-TM mode conversion. 
To isolate the intrinsic waveguide transmission from converter effects, and 
to rule out parasitic coupling between the two face-to-face converters, we 
performed de-embedding using a second, longer waveguide (190~mm, 10 isolated 
islands, plus converters). The transfer matrix $T$ of each fabricated device 
is written as a cascade of the input converter ($T_{ci}$), SSPP interconnect 
($T_{sp}$), and output converter ($T_{co}$):
\begin{equation}
    T_a = T_{ci}T_{sp,a}T_{co}, \qquad T_b = T_{ci}T_{sp,b}T_{co}
\end{equation}
Since the input and output converters are identical but flipped relative to 
each other, their matrices are related as
\begin{equation}
T_{ci} = \begin{bmatrix} T_{11,c} & T_{12,c} \\ T_{21,c} & T_{22,c} 
\end{bmatrix}, \quad
T_{co} = \begin{bmatrix} T_{11,c} & -T_{21,c} \\ -T_{12,c} & T_{22,c} 
\end{bmatrix}
\end{equation}
with $T_{11,c}T_{22,c}-T_{21,c}T_{12,c}=1$. The two SSPP waveguides 
$(sp,a)$ and $(sp,b)$ are uniform and identical except that the former is 
$n$ times longer, so $T_{sp,a} = (T_{sp,b})^n$. From the measured 
S-parameters of both devices and the converter relations above, we de-embed 
the transmission coefficient of the isolated-island SSPP interconnect.

Fig.~\ref{fig:experimental}(d) shows the raw and de-embedded $S_{21}$, the 
latter post-processed to remove measurement-sensitivity spikes. The 
de-embedded $S_{21}$ reaches a maximum of $-1$~dB over the 165~mm 
waveguide length, closely matching the numerically predicted $-3.1$~dB/m 
propagation loss. The majority of the loss seen in the raw measurement 
originates from inefficient TEM-to-SSPP mode conversion, identifying 
converter design as the primary target for future improvement.

The room-temperature measurements validate the electromagnetic design, 
establishing that signal propagation is governed by the guided-mode symmetry 
and is insensitive to the structural discontinuities introduced for thermal 
isolation. The thermal performance is characterized independently through the 
multiphysics simulation of Section~\ref{sec:thermal}, which also confirms 
that an applied thermal gradient shifts the SSPP cut-off frequency by 
$\sim$0.2~GHz.

\section{Conclusion}

This work demonstrates that the electromagnetic signal path and the thermal conduction path of a metallic guided-wave structure can be structurally decoupled through the symmetry of the supported mode. The odd-symmetric eigenmode of a double-edge-corrugated SSPP strip imposes a field null at the waveguide centerline, allowing the waveguide to be a periodic array of metallic islands. S-parameter measurements confirm signal propagation along 5-6.5 GHz window for the fabricated sample. Thermal analyses show $\sim$22$\times$ increase in thermal resistance when used periodic island array instead of standard microstrip lines. The resistance becomes $\sim$40$\times$ greater when adiabatic substrate voids are combined with isolated islands, resulting in an effective thermal conductivity of approximately 0.336 W/m$\cdot$K. 

Time-domain analysis further verified the interconnect’s capability to transmit AM modulated data pulses up to 1 Gbps data rate with high signal integrity. Furthermore, the inherent band-selective nature of the SSPP structure provides intrinsic filtering, while the operating frequency and bandwidth are tunable by geometric scaling of the corrugation depth, with no change to the fabrication process. The structure is planar, compatible with standard printed-circuit fabrication, and requires no precision mechanical alignment. It introduces a guided-wave design principle, electromagnetic transmission through a physically fragmented metallic path with suppressed thermal conduction, that is applicable wherever a microwave signal path must simultaneously route a signal and present a high thermal resistance along the propagation direction.


\section*{Acknowledgment}

The research is supported by University of North Carolina at Charlotte Faculty Startup Grant. The authors acknowledge support of the Microwave Laboratory of the Department of Electrical and Electronic Engineering, Bangladesh University of Engineering and Technology.


%





\ifCLASSOPTIONcaptionsoff
  \newpage
\fi





\bibliographystyle{IEEEtran}
\bibliography{IEEEabrv,Bibliography}

\vfill


\end{document}